# A Hierarchical Allometric Scaling Analysis of Chinese Cities: 1991-2014

Yanguang Chen, Jian Feng*(corresponding author)

(Department of Geography, College of Urban and Environmental Sciences, Peking University, Beijing 100871, P.R. China. E-mail: chenyg@pku.edu.cn; *fengjian@pku.edu.cn)

**Abstract**: The law of allometric scaling based on Zipf distributions can be employed to research hierarchies of cities in a geographical region. However, the allometric patterns are easily influenced by random disturbance from the noises in observational data. In theory, both the allometric growth law and Zipf's law are related to the hierarchical scaling laws associated with fractal structure. In this paper, the scaling laws of hierarchies with cascade structure are used to study Chinese cities, and the method of R/S analysis is applied to analyzing the change trend of the allometric scaling exponents. The results show that the hierarchical scaling relations of Chinese cities became clearer and clearer from 1991 to 2014 year; the global allometric scaling exponent values fluctuated around 0.85, and the local scaling exponent approached to 0.85. The Hurst exponent of the allometric parameter change is greater than 0.5, indicating persistence and a long-term memory of urban evolution. The main conclusions can be reached as follows: the allometric scaling law of cities represents an evolutionary order rather than an invariable rule, which emerges from self-organized process of urbanization, and the ideas from allometry and fractals can be combined to optimize spatial and hierarchical structure of urban systems in future city planning.

# 1. Introduction

Cities as systems (individuals) and systems of cities (groups) are scale-free complex systems,

which cannot be effectively described by the traditional mathematical methods based on characteristic scales in many respects. The ideas from scaling can be used to research urban systems (see e.g. Batty and Longley, 1994; Bettencourt, 2013; Bettencourt, 2007; Chen, 2008; Frankhauser, 1994; Lobo *et al*, 2013; Rybski *et al*, 2009). Two correlated scaling laws are often employed to analyze a hierarchy of cities: one is Zipf's law, and the other is the law of allometric growth. Zipf's law, allometric growth law, and distance-decay law compose three basic laws of urban geography. Zipf's law indicates the rank-size pattern of cities in a geographical region (Carroll, 1982; Gabaix, 1999a; Gabaix and Ioannides, 2004; Krugman, 1996; Zipf, 1949), and the allometric growth law describes the relationship between size and shape in the growth of human settlements (Batty, 2008; Dutton, 1973; Nordbeck, 1971; Lee, 1989; Lo and Welch, 1977). In fact, Zipf's law reflects urban growth (Batty and Longley, 1994; Gabaix, 1999a), and the allometric scaling law can be derived from dual Zipf's models of rank-size distributions of urban population and area (Chen, 2014a). This suggests that Zipf's law and the allometric scaling law represent different sides of the same coin. Zipf's law proved to be equivalent to a hierarchical scaling law, and the rank-size allometric scaling can be replaced by the hierarchical allometric scaling (Chen, 2012a). Hierarchy mirrors a universal structure in natural and social systems (Pumain, 2006). Zipf's law is a signature of hierarchical structure. Based on hierarchical scaling, we can develop a new approach to studying urban systems.

China bears a large set of cities with a long history. Studies on Chinese cities will help us understand the hidden order of complex systems. There is a dispute about whether the size distribution of Chinese cities follows Zipf's law (Anderson and Ge, 2005; Benguigui and Blumenfeld-Lieberthal, 2007a; Benguigui and Blumenfeld-Lieberthal, 2007b; Chen *et al*, 1993; Gangopadhyay and Basu, 2009; Ye and Xie, 2012). In fact, Zipf's law is a rule of evolution rather than that of existence. The rank-size pattern emerges from the edge of chaos (Bak, 1996). A new discovery is that Chinese city-size distribution can be described by the three-parameter Zipf's model instead of the two-parameter Zipf's model (Chen, 2016). The three-parameter Zipf's law suggests an incomplete hierarchy with cascade structure. This implies that the hierarchical scaling law can be used to research the allometry and rank-size pattern of Chinese cities.

This paper is devoted to making a hierarchical allomatric analysis of Chinese systems of cities. The aim of this study is as follows. First, we try to reveal the evolutional process and characteristics of the allometric scaling in Chinese cities. Second, we attempt to bring to light the causality behind the hierarchical allometry of Chinese cities. Third, we will sum up a general framework of hierarchical allometric analysis of cities. The trait of this case study rests with large samples, continuous time series, and new angle of view. By this work, we can obtain useful geographical information about spatio-temporal evolution of Chinese cities and new knowledge about scaling in cities. Moreover, the study lends further support to the suggestions that the geographical laws are evolutional laws and there are inherent relationships between Zipf's law and the law of allometric growth of cities. The rest parts of the paper are organized as below. In Section 2, the basic mathematical models of hierarchical structure are presented and explained; In Section 3, an empirical analysis of hierarchical allometric scaling in Chinese cities are made by means of two algorithms, to show the evolutional regularity of Chinese cities; In Section 4, several questions are discussed, and a general process of allometric analysis is proposed for urban studies. Finally, in Section 5, the article will be concluded by summarizing the mains of this study.

## 2. Models

### 2.1 Hierarchical scaling law of cascade structure

The mathematical models of self-similar hierarchies of cities can be expressed as a set of exponential functions and power functions. Using these models, we can make allometric scaling analysis based on city-size distribution. Suppose that the cities in a geographical region are grouped into $M$ classes in a top-down order according to the generalized $2^n$ principle (Chen, 2012a; Davis, 1978; Jiang and Yao, 2010). The cascade structure of the urban system can be modeled by three exponential equations as follows

$$f_m = f_1 r_f^{m-1}, \tag{1}$$

$$P_m = P_1 r_p^{1-m}, \tag{2}$$

$$A_m = A_1 r_a^{1-m}, \tag{3}$$

where $m$ refers to the top-down ordinal number of city level ($m$=1, 2, ⋯, $M$), $f_m$ denotes the number of cities of order $m$, correspondingly, $P_m$ and $A_m$ represent the mean population size and urban area at the $m$th level. The meaning of the parameters is as below: $f_1$ refers to the number of the top-level cities, $P_1$ and $A_1$ are the mean population size and urban area of the top-level cities, $r_f=f_{m+1}/f_m$ is the interclass **number ratio** of cities, $r_p=P_m/P_{m+1}$ is the population **size ratio**, and $r_a=A_m/A_{m+1}$ is the urban **area ratio**. Equations (1), (2) and (3) compose the mathematical expressions of the generalized $2^n$ rule (Chen, 2012a; Chen, 2012b), which is based on Beckmann-Davis models (Beckmann, 1958; Davis, 1978). In theory, if $r_f=2$ as given, then it will follow that $r_p \to 2$, and *vice versa*. Here the arrow denotes "approach" or "be close to". If $r_f= r_p=2$, the generalized $2^n$ rule will return to the normal $2^n$ rule presented (Jiang and Yao, 2010).

## 2.2 Hierarchical allometric rescaling

The cascade structure of a hierarchy of cities suggests allometry, fractal, and scaling. A set of power-law relations including the three-parameter Zipf's law can be derived from the above exponential laws (Chen, 2012b; Chen, 2016). The power-law models are as below:

$$f_m = \mu P_m^{-D}, \tag{4}$$

$$f_m = \eta A_m^{-d}, \tag{5}$$

$$A_m = a P_m^{b}, \tag{6}$$

where $\mu=f_1 P_1^D$, $D=\ln r_f/\ln r_p$, $\eta=f_1 A_1^d$, $d=\ln r_f/\ln r_a$, $a=A_1 P_1^{-b}$, and $b=\ln r_a/\ln r_p$. Equation (4) is termed the hierarchical *size-number scaling relation* of cities, which is equivalent to the Pareto law of city-size distribution, and $D$ is the fractal dimension of urban hierarchies measured with urban population. Equation (5) is termed the hierarchical *area-number scaling relation* of cities, which is equivalent to the Pareto law of city-area distribution, and $d$ can be treated as the fractal dimension of urban hierarchies measured with urban area. According to Chen (2012a), both $D$ and $d$ are actually paradimension rather than real fractal dimension. Equation (6) is termed the hierarchical allometric scaling relation between urban area and population, and $b$ is the allometric scaling exponent of an urban hierarchy. The inverse functions of equations (4) and (5) are equivalent to the Zipf's models of urban population size and urban area distributions. The allometric scaling exponent is actually the ratio of the fractal dimension of urban area size distribution to the fractal dimension of urban population size distribution, that is, $b=D/d$. Generally speaking, $d>D$, and $D$ is close to 1. Thus, the $b$ value comes between 0 and 1.

The exponential laws and the power laws reflect two relations of a hierarchy, respectively: longitudinal relations and latitudinal relations. The longitudinal relations are the associations across different classes, while the latitudinal relations are the correspondences between different measures such as city population size and urbanized area (Chen, 2012b). Formally, these hierarchical and measurement relations can be illustrated by dual hierarchies (Figure 1). If cities in a region follow Zipf's law, they can be organized into a hierarchy with cascade structure. On the other hand, if an urban system bears cascade structure, the cities in the system follow Zipf's law or can be described with a three-parameter Zipf's model. The Zipf distribution is a signature of self-similar hierarchies associated with fractal patterns and self-organized processes.

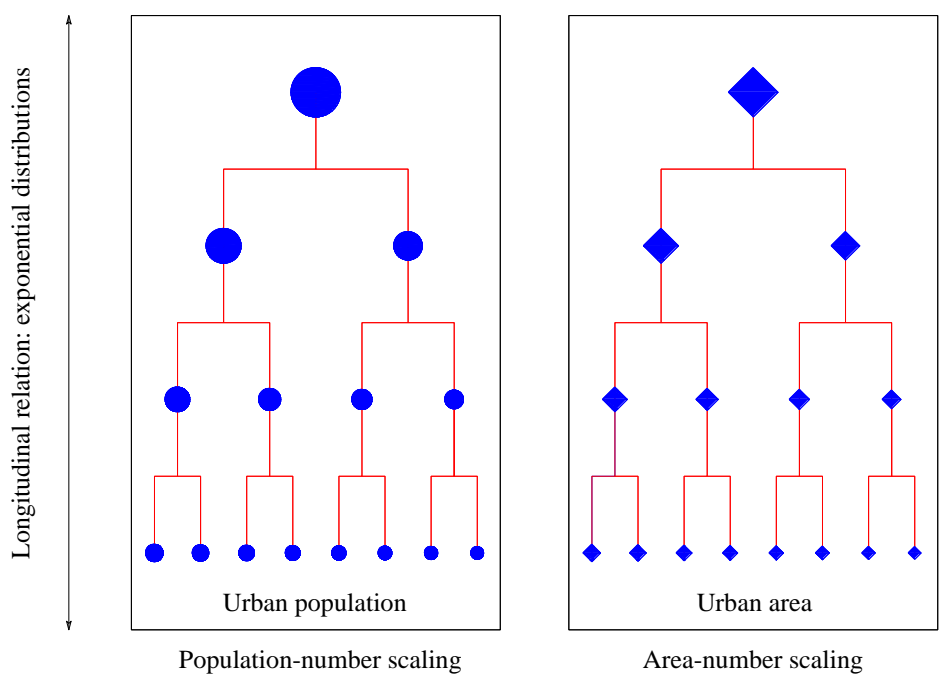

**Figure 1 A schematic diagram on the longitudinal distributions and latitudinal allometry of an urban hierarchy measured by population size and urban area (the first four classes)** [**Note:** See the appendices of Chen, 2012b]

## 3. Empirical analysis

### 3.1 Study area and methods

The models of hierarchical allometry can be applied to the rank-size distribution of Chinese cities. The study area contains the whole mainland of China, in which there are about 660 officially approved cities. The city number is different in different years. The city size is measured by the urban population within urbanized area, while the urban area is represented by the area of built district. A number of datasets of city sizes and urban area are available for this research, including the statistical data from 1991 to 2014 (24 years). All the observational data of city population size and urban area come from the Ministry of Housing and Urban-Rural Development of the People's Republic of China (MOHURD).

A number of algorithms can be employed to estimate the power exponent values, including the least squares method (LSM), maximum likelihood method (MLM), major axis method (MAM). Because of the advocacy of Newman (2005) and Clauset *et al* (2009), the MLM is treated as the only standard approach to fitting power law to observational data. In fact, the precondition of effectively making use of the MLM is that the observational data meet the joint normal distribution. However, for many social and economic systems such as cities, the observational variables do not satisfy the joint normal distribution. In this case, the LSM has its merits. The analytical process is as follows. **Step 1**, *preliminary analysis*. Using Zipf's law and the allometric growth law, we can examine the primary observational data sets based on urban population and area measurement. **Step 2**, *hierarchical reconstruction*. If the city-size or urban area distribution follows Zipf's law, and the relationships between urban area and population size follow the allometric scaling law, we can reconstruct the hierarchy by organizing the cities according to the model of cascade structure, equation (1). **Step 3**, *cascade analysis*. Using the exponential laws, equations (2) and (3), we can investigate the cascade structure of system of cities. **Step 4**, *allometric scaling analysis*. Using the power laws, equations (4), (5), and (6), we can research the hierarchical allometry of cities by urban size and area. Among various equations, equation (6) plays the most important role in our case study.

### 3.2 Results and findings

Now, let's make an allometric scaling analysis of hierarchy of Chinese cities step by step. First of all, we should examine the rank-size distribution of cities which can be described by Zipf's formulation. Zipf's law is universal rule followed by many natural and social systems (see e.g. Axtell, 2001; Bak, 1996; Batty, 2006; Gabaix, 1999a; Gabaix, 1999b; Hernando *et al*, 2009; Hong *et al*, 2007; Jiang and Jia, 2011; Shao *et al*, 2011; Stanley *et al*, 1995). This suggests that urban evolution follows the general laws of nature under certain condition. Zipf's distribution proved to be a signature of hierarchy with cascade structure (Chen, 2012b). If Chinese cities follow Zipf's law, they can be organized into a self-similar hierarchy. The two-parameter Zipf models of urban population and area can be expressed as

$$P_k = P_1 k^{-q}, \tag{7}$$

$$A_k = A_1 k^{-p}, \tag{8}$$

where $k$ denotes the rank of a city, $P_k$ and $A_k$ refer to the population size and built-up area of the $k$th city, $P_1$ and $A_1$ are the population size and urban area of the largest city, $q$ and $p$ are the Zipf scaling exponents. Using the OLS calculation, we can fit equations (7) and (8) to the observational data of urban population and area of Chinese cities. For example, where population size is concerned, in 2000, there were 665 officially approved cities, but only the largest 250 cities conform to Zipf's law; in 2010, however, there were 656 officially approved cities, but the largest 550 cities comply with Zipf's law approximately (Figure 2).

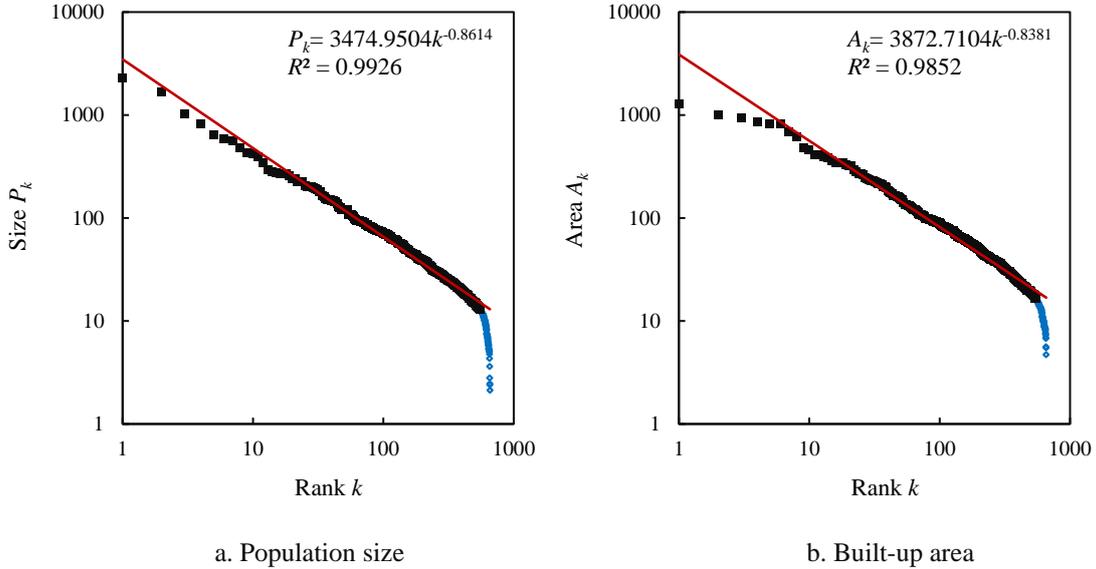

a. Population size    b. Built-up area

**Figure 2 The rank-size distributions of Chinese cities measured by urban population and built-up area in 2010**

Then, we investigate the allometric scaling relation based on the rank-size distribution. This is a kind of transversal allometry based on cross-sectional data (Pumain and Moriconi-Ebrard, 1997; Woldenberg, 1973). In fact, from equations (7) and (8) it follows

$$A_k = a P_k^b, \tag{9}$$

in which $a = A_1 P_1^{-b}$, $b = p/q = D/d$, where $D = 1/q$, $d = 1/p$ (Chen, 2014a). Fitting equations (9) to the observational data of population size and built-up area of Chinese cities yields models of cross-sectional allometry (Figure 3). In 2010, the relationship between urban population and area follow the law of allometric scaling. However, in 2000, the urban population and area takes on a linear relationship rather than a power-law pattern. In other word, in earlier years, Chinese cities departed from the allometric scaling law to some extent. This indicates that the scaling law of cross-sectional allometry is indeed associated with Zipf's law. Since Chinese cities failed to comply with Zipf's law in early years, the allometric scaling relation could not form. What is more, this lends further support to the suggestion that the laws of human geography are evolutional laws instead of existent laws (Chen, 2008; Feng and Chen, 2010).

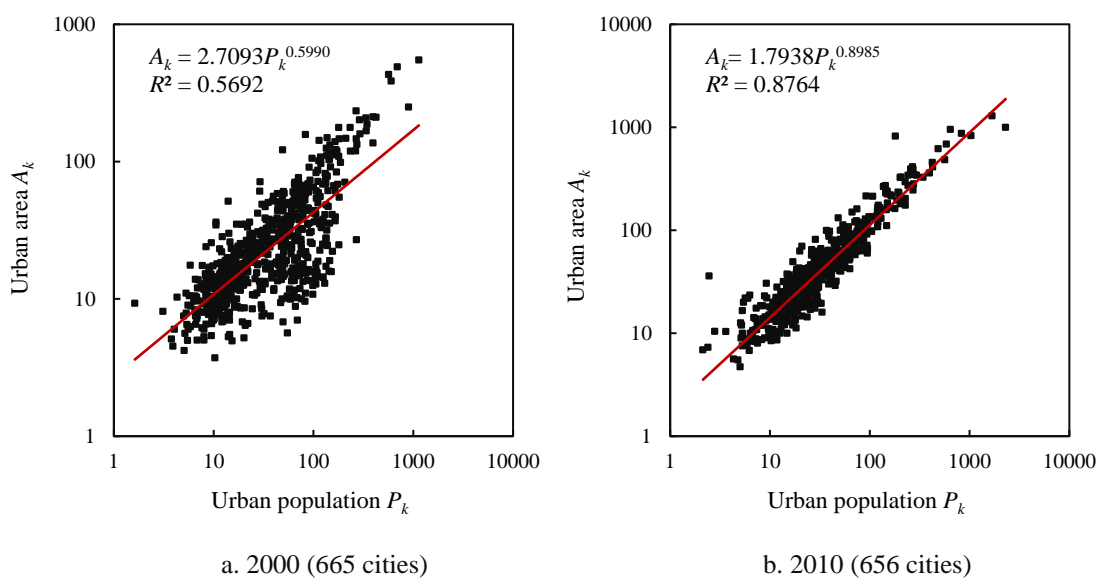

a. 2000 (665 cities)  b. 2010 (656 cities)

**Figure 3 The allometric scaling relationships between urban population and urbanized area based on the rank-size distributions of Chinese cities** [**Note**: Due to that some satellite towns were incorporated into vicinal large cities, the total number of officially approved cities fell from 2000 to 2010. In 2000, Chinese cities failed to follow the allometric scaling law.]

In the third step, we can check the cascade structure of urban hierarchy. The self-similar hierarchy can be constructed as follows. Suppose that we divide all the cities into $M$ classes/levels in a top-down order according to the generalized $2^n$ rule (Chen, 2008; Chen, 2012a&b; Jiang and Yao, 2010). The result is one city in the first class (rank 1), two cities in the second class (ranks 2 and 3), four cities in the third class (ranks 4, 5, 6 and 7), and so on (Figure 1). This cascade structure is based on city population rather than urban area. Generally speaking, urban area development lags behind population size growth of a city. Then, we can calculate the average population size and average urban area of the cities at different levels (Table 1). If the average population size and average urban area follow the exponential laws, we can describe the size and area decay using equations (2) and (3). The results show that the average population size and area take on exponent decay patterns (Figure 4). This suggests that the relationships between average urban population and area in different classes follows the power laws and can be described by equations (4), (5), and (6). The results show that the first and the last classes are sometimes exceptional points, and the other points form a scaling range and comply with the power laws (Figure 5). The scaling exponent of equation (4) gives the fractal dimension of population size distribution ($D=1/q$), while the scaling exponent of equation (5) yields the fractal dimension of urban area distribution ($d=1/p$) (Chen, 2012a).

**Table 1 The examples of reconstructed self-similar hierarchies of Chinese cities in 2000 and 2010**

| Class | 2000 | | | 2010 | | |
|---|---|---|---|---|---|---|
| $m$ | City number $f_m$ | Average size $P_m$ | Average area $A_m$ | City number $f_m$ | Average size $P_m$ | Average area $A_m$ |
| 1 | 1 | 1136.8200 | 549.5800 | 1 | 2301.9100 | 998.7500 |
| 2 | 2 | 793.6650 | 369.8250 | 2 | 1360.8450 | 1059.6550 |
| 3 | 4 | 496.3750 | 290.7500 | 4 | 657.1000 | 748.2450 |
| 4 | 8 | 317.3288 | 161.8475 | 8 | 366.0988 | 392.9913 |
| 5 | 16 | 205.6356 | 118.5538 | 16 | 226.9919 | 276.7556 |
| 6 | 32 | 146.9441 | 67.7275 | 32 | 129.5641 | 180.1488 |
| 7 | 64 | 107.9066 | 44.9964 | 64 | 74.8284 | 88.7334 |
| 8 | 128 | 65.8054 | 34.6786 | 128 | 41.4015 | 53.6396 |
| 9 | 256 | 26.5613 | 21.5523 | 256 | 21.1980 | 27.7446 |
| 10 | 154 | 9.4431 | 11.5560 | 145 | 10.3430 | 15.5314 |

**Note**: These results are similar to the results based on urban census data, which were displayed in Chen (2016), but there are subtle difference. Unfortunately, we have no datasets for urbanized area of Chinese cities.

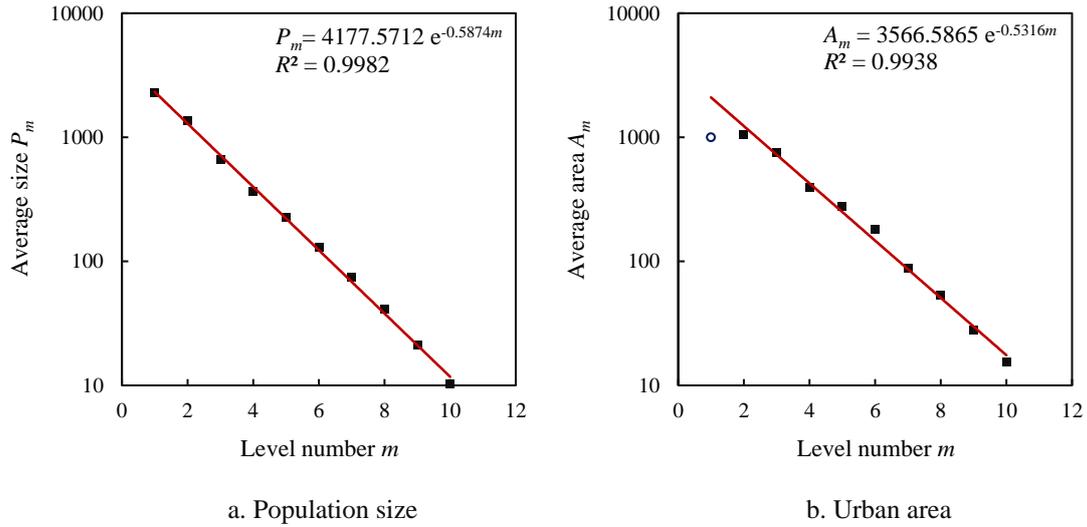

a. Population size  b. Urban area

**Figure 4 The patterns of exponential decay of population size and built-up area of Chinese cities in 2010** [**Note:** The small circle represents the outlier indicating the top-level cities.]

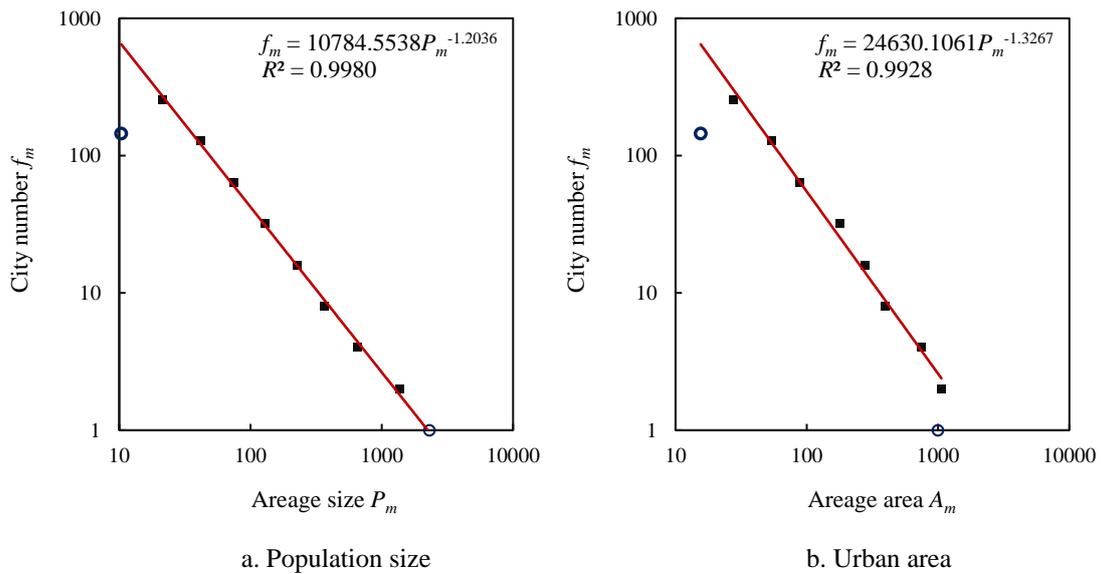

a. Population size  b. Urban area

**Figure 5 The hierarchical scaling relationships between population size, built-up area, and number of Chinese cities in 2010** [**Note:** The small circles represent top classes and the lame-duck classes. The slopes based on the scaling ranges indicate the fractal parameters of city size and area distributions. The ratio of the size dimension $D$ to the area dimension $d$ is an approximate value of the allometric scaling exponent $b$.]

In the fourth step, we should research the hierarchical allometric scaling relationships between urban population and area. This step is the focus of this study. By statistical analysis and double logarithm plots, we can discover a number of new phenomena and trends (Figure 6). ***First, the*** ***allometric scaling relation emerges from the self-organized process of urbanization.*** From 1991 to 1992 year, the urban population-area relation follows the power law indicative allometric scaling, but the (local) allometric exponent is greater than 1. The proper allometric scaling exponent should come between 2/3 and 1 (Table 2). From 1993 to 2002 year, the local allometric relation based on scaling range degenerated to a linear relation. In particular, during the period from 1998 to 2000, both the global allometry and local allometry degenerated to linear relations (Table 3). In other words, it is the linear function rather than the power function that can be best fitted to the observational data of 1998, 1999, and 2000. This indicates that the allometric scaling relation between urban population and area has not yet emerged before 2003. The power law relation can be constrainedly fitted to the observational data from 1993 to 2002 year, but the points do not well match the trend lines. In fact, at the early stage, the medium-sized cities with population from 50 to 150 thousands did not develop where urbanized area is concerned. From 2003 on, the allometric scaling relation emerged, and this relation became significant in 2004. Once again, this lends further support to the suggestion that geographical laws are evolutionary laws instead of everlasting laws. What is more, the allometric scaling exponent ranged from 2/3 to 1. However, from 2002 on, the land of the outsize cities with population greater than 400 thousands was overused because of real estate bubble economy. ***Second, the allometric scaling law depends on Zipf's law.*** Generally speaking, population growth precedes the development of urban area indicative of land use, but the rank-size distribution of urban area is more stable than that of population. However, for Chinese cities, the thing is abnormal. For example, from 1998 to 2000, the 500 largest Chinese cities followed Zipf's law of urban area, but only about 250 largest cities complied with the common Zipf's law of city population. In other words, the rank-population distribution was not consistent with the rank-area distribution. As a result, the allometric scaling relation of Chinese cities degenerated from power law relation to quasi-linear relation. This lends further support to the inference that urban area-population allometry proceeds from the scaling relation between the Zipf's distribution of city population and that of urban area (Chen, 2014a). ***Third, on the whole, the allometric scaling exponent value changed around constant***. In theory, the scaling exponent of allometric growth ($b$) is the ratio of the fractal dimension of urban land-use form ($D_f$) to that of urban population distribution ($D_p$) (Chen, 2014a). Thus, the allometric exponent is supposed to fluctuates between 2/3 and 1 and approaches 0.85 because $D_f \rightarrow 1.7$ and $D_p \rightarrow 2$ (Chen, 2010).

However, in practice, the computation yields a fraction sequence such as 1/2, 2/3, 3/4, …, $n/(n+1)$,…, where *n* refers to the set of natural number (Chen, 2014b). ***Fourth, there is scale-free range in which the allometric relation is more significant***. On log-log plots, two data points often take on outliers: one is first class, and the other is the last class. On the one hand, number one is always a special one in theory (Chen, 2012a). The top class indicating the largest city is usually different from other classes and departs from the trend line. On the other, the bottom class also manifests an exceptional value in a plot due to undergrowth of small cities or absence of partial data (Chen, 2016). The exceptional values are clear in the plots of rank-size distributions and the corresponding size-number scaling relation (see Figure 5). The other points coming between the first and last classes form a straight line and represent a scaling range. However, the outliers of rank-size distribution or size-number scaling often become insignificant in the hierarchical allometric relation (Figure 6). Despite this, if we fit the model of allometric scaling to the data points within the scaling range, the allometric exponent values will rise, and the mean is about 0.95 (Table 2). Please note that the *local allometry* is based on the data points within the scaling range in this context. Accordingly, the *global allometry* is based on all the data points, including the data points inside and outside the scaling range.

**Table 2 The global and local allometric and fractal parameters of the hierarchies of Chinese cities (1991-2014)**

| Year | Results based on all the data points (global allometry) | | | | | | Results based on the scaling range (local allometry) | | | | | |
|---|---|---|---|---|---|---|---|---|---|---|---|---|
| | *a* | *b* | $R^2$ | *D* | *d* | *D/d* | *a* | $b^*$ | $R^2$ | $D^*$ | $d^*$ | $D^*/d^*$ |
| **1991** | 0.7609 | 0.8977 | 0.9479 | 1.5724 | 1.7040 | 0.9228 | 0.2321 | 1.1228 | 0.9727 | 1.8018 | 1.5664 | 1.1503 |
| **1992** | 0.7818 | 0.8960 | 0.9451 | 1.5947 | 1.7222 | 0.9260 | 0.2171 | 1.1388 | 0.9763 | 1.8120 | 1.5555 | 1.1649 |
| **1993** | 1.6289 | 0.7622 | 0.9442 | 0.9320 | 1.3215 | 0.7053 | 0.6375 | 0.9396 | 0.9554 | 1.5964 | 1.6677 | 0.9572 |
| **1994** | 1.4179 | 0.7904 | 0.9533 | 1.1120 | 1.4537 | 0.7649 | 0.6310 | 0.9439 | 0.9527 | 1.6181 | 1.6838 | 0.9610 |
| **1995** | 1.3064 | 0.8123 | 0.9468 | 1.1809 | 1.4848 | 0.7953 | 0.5090 | 0.9931 | 0.9497 | 1.6568 | 1.6832 | 0.9843 |
| **1996** | 1.2484 | 0.8383 | 0.9566 | 1.2164 | 1.4839 | 0.8197 | 0.6597 | 0.9554 | 0.9487 | 1.5976 | 1.6436 | 0.9720 |
| **1997** | 1.2916 | 0.8334 | 0.9555 | 1.2275 | 1.5020 | 0.8172 | 0.6622 | 0.9568 | 0.9484 | 1.5995 | 1.6437 | 0.9731 |
| **1998** | 1.2289 | 0.8483 | 0.9666 | 1.2078 | 1.4593 | 0.8277 | 0.7932 | 0.9217 | 0.9611 | 1.5462 | 1.6630 | 0.9298 |
| **1999** | 1.2430 | 0.8458 | 0.9693 | 1.2066 | 1.4610 | 0.8259 | 0.7934 | 0.9225 | 0.9643 | 1.5426 | 1.6595 | 0.9296 |
| **2000** | 1.2885 | 0.8396 | 0.9721 | 1.2046 | 1.4674 | 0.8209 | 0.8417 | 0.9114 | 0.9690 | 1.5231 | 1.6571 | 0.9191 |
| **2001** | 1.4402 | 0.8419 | 0.9869 | 1.1608 | 1.3963 | 0.8313 | 1.0042 | 0.9096 | 0.9860 | 1.3900 | 1.5293 | 0.9089 |
| **2002** | 1.4268 | 0.8633 | 0.9818 | 1.1352 | 1.3161 | 0.8625 | 0.9371 | 0.9470 | 0.9828 | 1.3395 | 1.4099 | 0.9501 |
| **2003** | 1.5490 | 0.8720 | 0.9755 | 1.1191 | 1.2667 | 0.8835 | 0.9883 | 0.9661 | 0.9822 | 1.2888 | 1.3292 | 0.9696 |

| Year | | | | | | | | | | | |
|---|---|---|---|---|---|---|---|---|---|---|---|
| **2004** | 1.5335 | 0.8859 | 0.9905 | 1.1198 | 1.2651 | 0.8851 | 1.2023 | 0.9333 | 0.9880 | 1.2754 | 1.3586 | 0.9388 |
| **2005** | 1.7848 | 0.8556 | 0.9879 | 1.0796 | 1.2489 | 0.8644 | 1.3683 | 0.9129 | 0.9879 | 1.2322 | 1.3400 | 0.9196 |
| **2006** | 1.8313 | 0.8713 | 0.9884 | 1.0717 | 1.2122 | 0.8841 | 1.3808 | 0.9369 | 0.9986 | 1.2171 | 1.2985 | 0.9373 |
| **2007** | 1.9359 | 0.8680 | 0.9871 | 1.0705 | 1.2115 | 0.8836 | 1.4771 | 0.9323 | 0.9982 | 1.2128 | 1.2990 | 0.9336 |
| **2008** | 2.0198 | 0.8687 | 0.9846 | 1.0752 | 1.2116 | 0.8874 | 1.5383 | 0.9346 | 0.9965 | 1.2225 | 1.3032 | 0.9381 |
| **2009** | 2.1126 | 0.8668 | 0.9833 | 1.0752 | 1.2136 | 0.8860 | 1.5698 | 0.9372 | 0.9954 | 1.2238 | 1.2994 | 0.9418 |
| **2010** | 2.4915 | 0.8359 | 0.9784 | 1.0552 | 1.2279 | 0.8594 | 1.9223 | 0.9012 | 0.9919 | 1.2036 | 1.3267 | 0.9072 |
| **2011** | 2.6215 | 0.8353 | 0.9715 | 1.0697 | 1.2386 | 0.8636 | 1.8691 | 0.9179 | 0.9898 | 1.2426 | 1.3421 | 0.9259 |
| **2012** | 2.7957 | 0.8228 | 0.9702 | 1.0542 | 1.2349 | 0.8537 | 2.1355 | 0.8910 | 0.9837 | 1.2105 | 1.3408 | 0.9028 |
| **2013** | 2.9567 | 0.8171 | 0.9675 | 1.0519 | 1.2362 | 0.8509 | 2.2495 | 0.8867 | 0.9823 | 1.2114 | 1.3467 | 0.8995 |
| **2014** | 2.9809 | 0.8194 | 0.9662 | 1.0489 | 1.2253 | 0.8560 | 2.2490 | 0.8913 | 0.9839 | 1.2099 | 1.3397 | 0.9031 |

**Note**: (1) The notations are as follows: $a$--proportionality coefficient, $b$--allometric scaling exponent, $R^2$-- goodness of fit, $D$--size dimension (the fractal dimension of population size distribution), $d$ --area dimension (the fractal dimension of urban area distribution), $D/d$ -- fractal dimension quotient or dimension ratio. (2) Removing the first class and the last class yields the scaling range.

**Table 3 The evolution of the allometric scaling in the hierarchy of Chinese cities (1991-2014)**

| Year | City number | Global analysis | Local analysis | Year | City number | Global analysis | Local analysis |
|---|---|---|---|---|---|---|---|
| 1991 | 473 | Power law | Power law | 2003 | 658 | Power law | Power law |
| 1992 | 506 | Power law | Power law | 2004 | 659 | Power law | Power law |
| 1993 | 551 | Power law | Linear relation | 2005 | 659 | Power law | Power law |
| 1994 | 610 | Power law | Linear relation | 2006 | 655 | Power law | Power law |
| 1995 | 633 | Power law | Linear relation | 2007 | 655 | Power law | Power law |
| 1996 | 658 | Power law | Linear relation | 2008 | 655 | Power law | Power law |
| 1997 | 669 | Power law | Linear relation | 2009 | 654 | Power law | Power law |
| 1998 | 667 | Linear relation | Linear relation | 2010 | 656 | Power law | Power law |
| 1999 | 665 | Linear relation | Linear relation | 2011 | 657 | Power law | Power law |
| 2000 | 665 | Linear relation | Linear relation | 2012 | 655 | Power law | Power law |
| 2001 | 664 | Power law | Linear relation | 2013 | 658 | Power law | Power law |
| 2002 | 660 | Power law | Linear relation | 2014 | 658 | Power law | Power law |

**Note**: The global analysis is based on all the classes, which can be represented by all the data points on the log-log plots. The local analysis can be made by removing the first class and the last class. For the power-law relation, the local analysis is based on the scaling range.

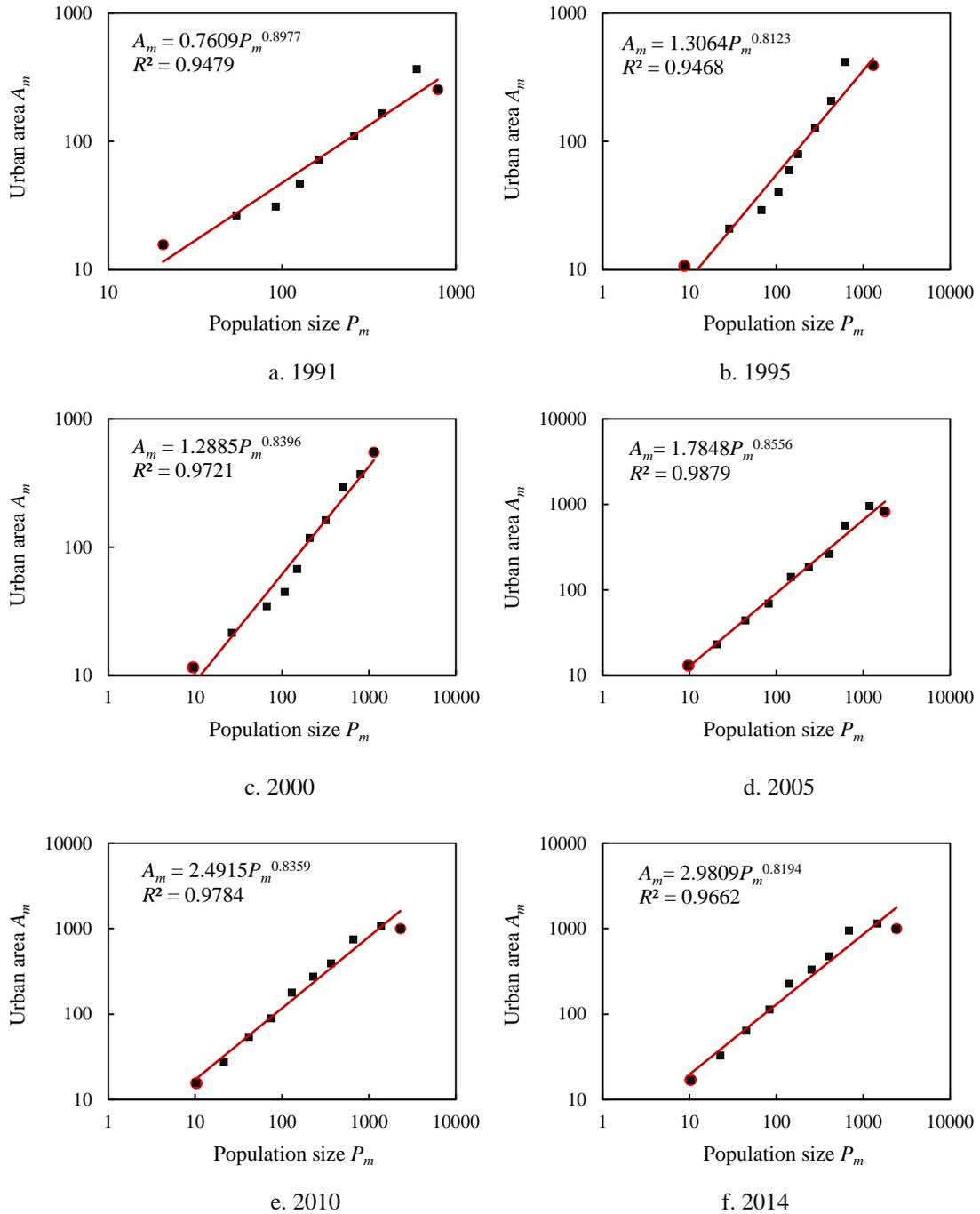

**Figure 6 The hierarchical allometric scaling relationships between population size and area of built district of Chinese cities (examples)**

[**Note:** The small circles represent the largest city at the top class and the bottom cities at the lame-duck classes. The trend lines are based on all the data points rather that those within the scaling ranges.]

## 3.3 The R/S analysis of allometric scaling exponents

The changing trend of the allometric parameter values reflects the trend of urban evolution. As shown above, four sets of parameters have been evaluated. The first is the allometric scaling exponent based on the global analysis, $b$; the second is the estimated value of $b$ by the ratio of the fractal dimension of city population size distribution (size dimension) to that of urban area distribution (area dimension) based on scaling ranges, $D/d$; the third is the allometric scaling exponent based on the local analysis, $b^*$; the fourth is the estimated value of $b^*$ by the ratio of the size dimension to area dimension based on scaling ranges, $D^*/d^*$. The global allometric exponent $b$ and its estimated result $D/d$ fluctuated around the mean value 0.85, but the local allometric exponent $b^*$ and its estimated result $D^*/d^*$ descended and takes on a trend of damped vibration—gradually decayed and approached to 0.85 (the empirical value is about 6/7).

The method of the *rescaled range analysis* can be employed to predict the future direction of parameter change, which indicates the trend of urban evolution. The rescaled range analysis is also termed *R/S* analysis, which is an approach of nonlinear time series analysis put forward by Hurst *et al* (1965). The basic parameter of the *R/S* analysis is Hurst exponent, which is associated with the self-affine fractal dimension and autocorrelation coefficient (Chen, 2013; Feder, 1988; Mandelbrot, 1982). The Hurst exponent value comes between 0 and 1. If the Hurst exponent is equal to 1/2, the time series bears no long memory and urban change has no autocorrelation; If the Hurst exponent is greater than 1/2 significantly, the time series bears persistence and urban change has positive autocorrelation; If the Hurst exponent is less than 1/2 significantly, the time series bears anti-persistence and urban change has negative autocorrelation. The calculations indicate that the method is suitable for time series analysis of the allometric scaling exponents (Figure 7). The results show that all the Hurst exponent values are greater than 1/2 (Table 4). This suggests that global allometric exponent will change around the mean value as usual, but the local allometric exponent will continue to decay before it arrives at the expected value, 0.85. An inference is that the global allometric exponent and the local allometric exponent will reach the same goal by different routes.

Table 4 The Hurst exponent values of the evolutional hierarchy of Chinese cities (1991-2014)

| Quantity | $b$ | $D/d$ | $b^*$ | $D^*/d^*$ |
| --- | --- | --- | --- | --- |
| **Proportionality Coefficient** | 1.0509 | 1.0958 | 1.1293 | 1.1429 |
| **Hurst exponent** | 0.6416 | 0.5961 | 0.5409 | 0.5523 |
| **Goodness of fit** | 0.9924 | 0.9943 | 0.9907 | 0.9884 |
| **Autocorrelation coefficient** | 0.2168 | 0.1426 | 0.0584 | 0.0752 |

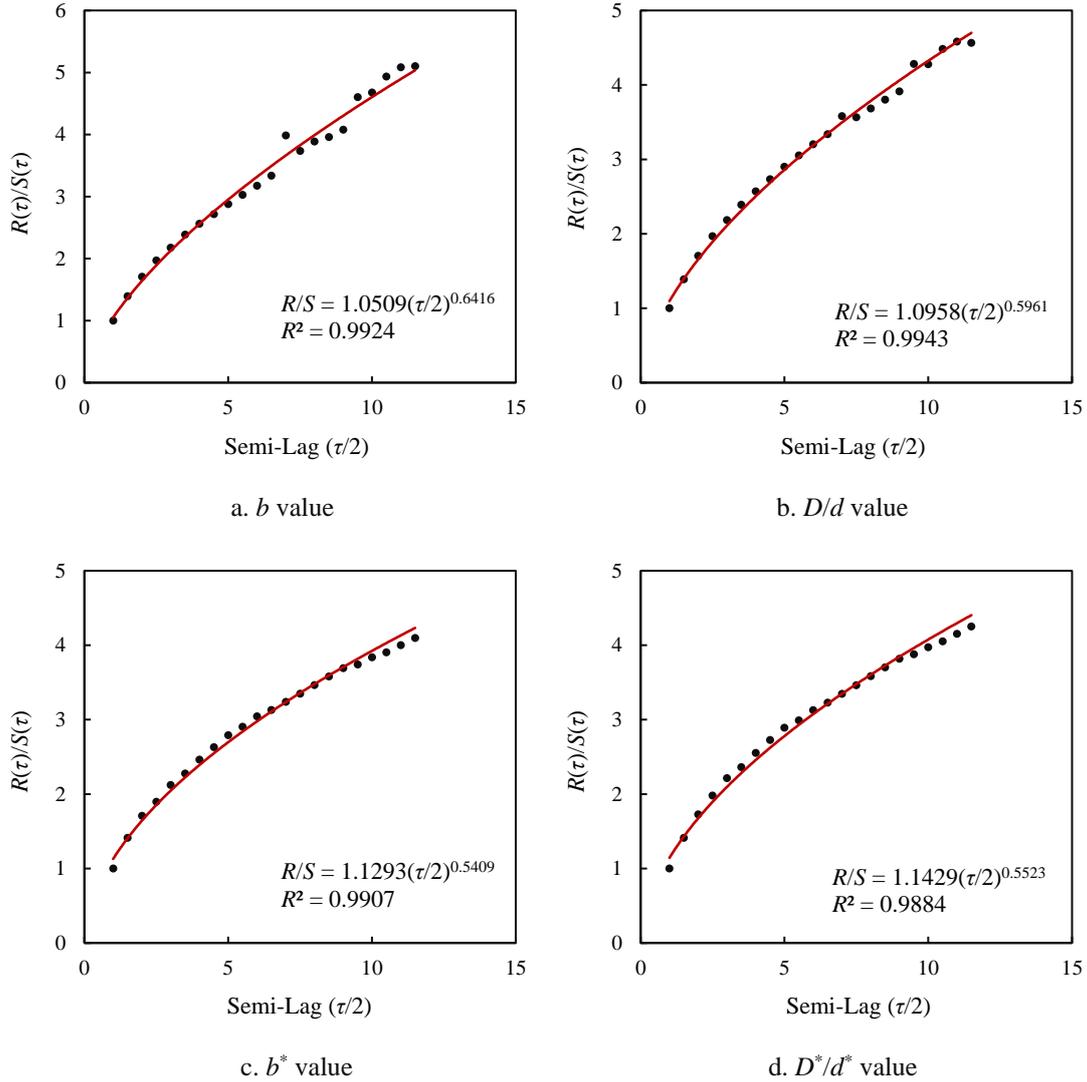

**Figure 7 The plots for R/S analysis of allometric scaling exponents of Chinese cities: 1991-2014**

### 3.4 The results based on maximum-likelihood fitting

The algorithm adopted in this study is in fact the ordinary least squares (OLS) method, which can be easily applied to double logarithmic linear regression. In fact, OLS is a conventional approach to estimating the power exponents. However, the OLS algorithm is not always the best one for fitting power laws to observational data. A new method based on the maximum likelihood estimation (MLE) is developed by Clauset *et al* (2009) to address the power-law distribution in empirical data. Combining the methods of maximum-likelihood fitting with the goodness-of-fit tests based on the Kolmogorov-Smirnov statistic and likelihood ratios, the approach is employed by many scholars to identify various power-law distributions in both natural and social sciences. Unfortunately, the MLE-based approach cannot be directly applied to our datasets in this work. The reasons are as follows. First, the MLE-based on method is developed for power-law frequency distributions, while this study is devoted to allometric scaling relations. Second, the MLE-based method is suitable for binned data, while this study is based on pairs of cascade sequences.

However, the MLE-based approach can be indirectly applied to the hierarchical allometric scaling analysis. As indicated above, an allometric scaling relation can be derived from two Zipf's distributions, and the hierarchical allometry is equivalent to the rank-size allometry (Chen, 2010; Chen, 2016). The MLE-based method developed by Clauset and his co-workers can be used to research Zipf's distributions. If urban population size distribution and area size distributions follow Zipf's law simultaneously, the relation between urban area and population will follow the allometric scaling law (Chen, 2008); If the urban area-population relation follows the rank-size allometric scaling law, it will follow the hierarchical allometric scaling law (Chen, 2014a). Thus, we can apply the MLE-based approach to the Zipf's distributions of urban population and area in different years, respectively. Zipf's distribution is mathematically equivalent to Pareto distribution (Chen *et al*, 1993). After the Pareto exponents are estimated, we can estimate the allometric scaling exponent based on maximum likelihood fitting. For a power-law distribution function $f(x) \propto x^{-\beta}$, the corresponding density distribution can be obtained by differential as follows

$$N(x) = \frac{\mathrm{d}f(x)}{\mathrm{d}x} \propto x^{-(\beta+1)} \propto x^{-\alpha}, \tag{10}$$

where $f(x)$ is cumulative distribution function (CDF), $N(x)$ is density distribution function (DDF), $\beta$ denotes the Pareto exponent, i.e., the cumulative scaling exponent, $\alpha=\beta+1$ represents the density scaling exponent, d refers to differential operator, and $\propto$ to the proportional relation. Replacing the general unknown quantity $x$ by urban population ($P$) and area ($A$) yields two density distribution functions as below

$$N(P) \propto P^{-\alpha_p} \propto P^{-(1+D)}, \tag{11}$$

$$N(A) \propto A^{-\alpha_a} \propto P^{-(1+d)}. \tag{12}$$

in which $\alpha_p$ and $\alpha_a$ represents the density scaling exponents of urban population and area distributions, and the other symbols are in essence the same as in equations (4) and (5). Thus the allometric scaling exponent can be estimated by the formula $b=D/d$, as shown above. For the purpose of comparison, both the MLE-based approach and the OLS-based approach are applied to the datasets of Zipf's distributions of Chinese urban population and area from 1991 to 2014. The estimated values of the allometric scaling exponents are tabulated as follows (Table 5).

Table 5 A comparison between the allometric scaling exponents based on MLE and those based on OLS

| Year | MLE-based results | | | | | OLS-based results | |
| --- | --- | --- | --- | --- | --- | --- | --- |
| | $α_p$ | $α_a$ | $D$ | $d$ | $b^*$ | $b^{**}$ | $R^2$ |
| **1991** | 3.3627 | 2.1953 | 2.3627 | 1.1953 | 1.9767 | 0.5277 | 0.3680 |
| **1992** | 3.1902 | 2.2092 | 2.1902 | 1.2092 | 1.8113 | 0.5115 | 0.3904 |
| **1993** | 3.2499 | 2.6439 | 2.2499 | 1.6439 | 1.3686 | 0.5065 | 0.3711 |
| **1994** | 3.3373 | 2.2475 | 2.3373 | 1.2475 | 1.8736 | 0.5132 | 0.4158 |
| **1995** | 3.5408 | 2.9049 | 2.5408 | 1.9049 | 1.3338 | 0.5324 | 0.4402 |
| **1996** | 3.3537 | 2.1760 | 2.3537 | 1.1760 | 2.0014 | 0.5631 | 0.5043 |
| **1997** | 3.2685 | 2.1876 | 2.2685 | 1.1876 | 1.9102 | 0.5679 | 0.5193 |
| **1998** | 3.3128 | 2.5258 | 2.3128 | 1.5258 | 1.5158 | 0.5873 | 0.5502 |
| **1999** | 2.1752 | 3.4990 | 1.1752 | 2.4990 | 0.4703 | 0.5908 | 0.5571 |
| **2000** | 3.1175 | 2.1752 | 2.1175 | 1.1752 | 1.8018 | 0.5990 | 0.5692 |
| **2001** | 2.7759 | 2.5755 | 1.7759 | 1.5755 | 1.1272 | 0.6834 | 0.7054 |
| **2002** | 2.9099 | 2.5192 | 1.9099 | 1.5192 | 1.2572 | 0.7328 | 0.7922 |
| **2003** | 1.9351 | 2.4149 | 0.9351 | 1.4149 | 0.6609 | 0.7837 | 0.8339 |
| **2004** | 1.9716 | 2.4939 | 0.9716 | 1.4939 | 0.6504 | 0.8036 | 0.8373 |
| **2005** | 1.9901 | 2.4060 | 0.9901 | 1.4060 | 0.7042 | 0.8160 | 0.8415 |
| **2006** | 2.0938 | 2.1734 | 1.0938 | 1.1734 | 0.9322 | 0.8616 | 0.8771 |
| **2007** | 2.3322 | 2.3922 | 1.3322 | 1.3922 | 0.9569 | 0.8798 | 0.8763 |
| **2008** | 2.1116 | 2.3188 | 1.1116 | 1.3188 | 0.8429 | 0.9029 | 0.8883 |
| **2009** | 2.3777 | 2.2339 | 1.3777 | 1.2339 | 1.1165 | 0.8977 | 0.8862 |
| **2010** | 2.3387 | 2.2373 | 1.3387 | 1.2373 | 1.0820 | 0.8985 | 0.8764 |
| **2011** | 2.1632 | 2.3573 | 1.1632 | 1.3573 | 0.8570 | 0.8954 | 0.8770 |
| **2012** | 2.1875 | 2.3560 | 1.1875 | 1.3560 | 0.8757 | 0.8923 | 0.8775 |
| **2013** | 2.1670 | 2.3210 | 1.1670 | 1.3210 | 0.8834 | 0.8644 | 0.8709 |
| **2014** | 2.1237 | 2.3043 | 1.1237 | 1.3043 | 0.8615 | 0.8741 | 0.8761 |

**Note:** The MLE-based allometric scaling exponent $b^*$ and the OLS-based allometric scaling exponent $b^{**}$ values are both based on Zipf's distributions, and are estimated by the rank-size allometric scaling relation. The $b^{**}$ values are different from the results shown in Table 2, which is based on hierarchical allometric scaling.

The MLE-based results lend further support to the findings shown in Subsection 3.2 (Results and findings). In theory, the reasonable density scaling exponent comes between 1.5 and 3, while the logical allometric scaling exponent varies from 2/3 to 1. However, from 1991 to 2002, the MLE-based allometric scaling exponent values are abnormal, ranging from 0.47 to 2. Accordingly, the OLS-based results are also abnormal, ranging from 0.5 to 0.74. There are significant difference between the MLE-based results and the OLS-based results (Table 5). From 2003 to 2014, the MLE-based results and the OLS-based results seem to reach the same goal by different routes (Figure 8). As discussed above, the year 2003 is a turning point. Before 2003, the allometric scaling was not significant. From 2003 on, the allometric scaling relation between urban population and area emerged. The findings are as follows. First, the MLE-based method developed by Clauset et al (2009) is rather more sensitive to the power-law relation than the OLS-based method. Second, the similarity and differences between the MLE-based results and the OLS-based results are helpful for us to examine the evolution of allometric scaling. Third, as a whole, the MLE-based method is not suitable for this studies, but the two methods can supplement each other.

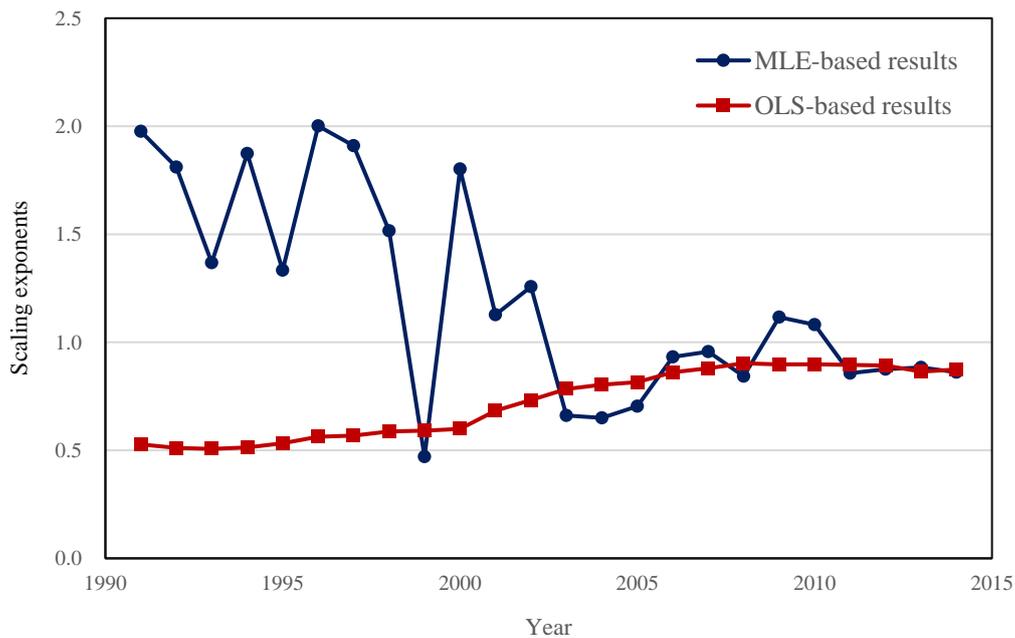

**Figure 8 A comparison between the change trend of MLE-based allometric parameter values and that of OLS-based parameter values [Note:** Before 2002, there was a significant difference between the MLE-based results and OLS-based results. From 2003 on, the two sets of results went to be consistent gradually.**]**

## 4. Discussion

Hierarchical scaling suggests a universal law of complex systems such as cities. The scaling law is followed by central place network (Chen, 2012b), river composition (Horton, 1945; Rodriguez-Iturbe and Rinaldo, 2001; Schumm, 1956; Strahler, 1952), earthquake energy distribution (Gutenberg and Richter, 1954; Turcotte, 1997), animals' blood vessels (Chen, 2015; Jiang and He, 1989; Jiang and He, 1990), and so on. Based on the case studies of hierarchy of Chinese cities, a complete process of allometric scaling analysis is developed. The traditional forms of Zipf's law and allometric growth law can be integrated into the new analytical framework (Figure 9). Using the hierarchical scaling method, we can research city development and urban evolution from a new angle of view. By the empirical analysis, we obtain an insight into Chinese urban system. **First, Zipf's law and the allometric growth law are not "iron laws" for Chinese cities.** The development of allometric scaling in Chinese cities can be divided into three stages. The first stage is the abnormal allometry phase (before 1992), during which the urban population area relation follow the power law, but this local allometric exponent exceeded the proper upper limit ($b>1$). The second stage is the allometry-breaking phase (1993-2002), during which the global allometric scaling or even the local allometric scaling degenerated to linear relations. The third stage is normal allometry phase (after 2003), during which the urban population-area relation follow the power law, and the allometric exponent come between 2/3 and 1 ($2/3<b<1$). **Second, the scaling relations are significantly influenced by fast urbanization and real estate industry of China.** Fast urbanization results in rapid growth of urban population, and realty industry leads to built-up area change hastily. The relationships between urbanization and real estate industry were imprinted on Chinese city development. During the period of fast urbanization, the real estate bubble led to many rounds of "creates city" campaigns in China. **Third, two types of forces impact on urban evolution of China.** One is the top-down force coming from command economy and government intervention, and the other is the bottom-up force coming from market economy and individual actions. The former suggests the Invisible Hand, and the latter indicates the Visible Hand. In this case, the allometric scaling and rank-size patterns of Chinese cities seem to struggle between order and chaos. Despite all these, the allometric scaling exponent is very stable in the main. In fact, a scaling exponent is often close to a ratio such as 1/2, 2/3, 3/4, 4/5, 5/6, and so forth (Chen, 2014). The allometric exponent values of Chinese cities seemed to change around 6/7 from 1991 to 2014.

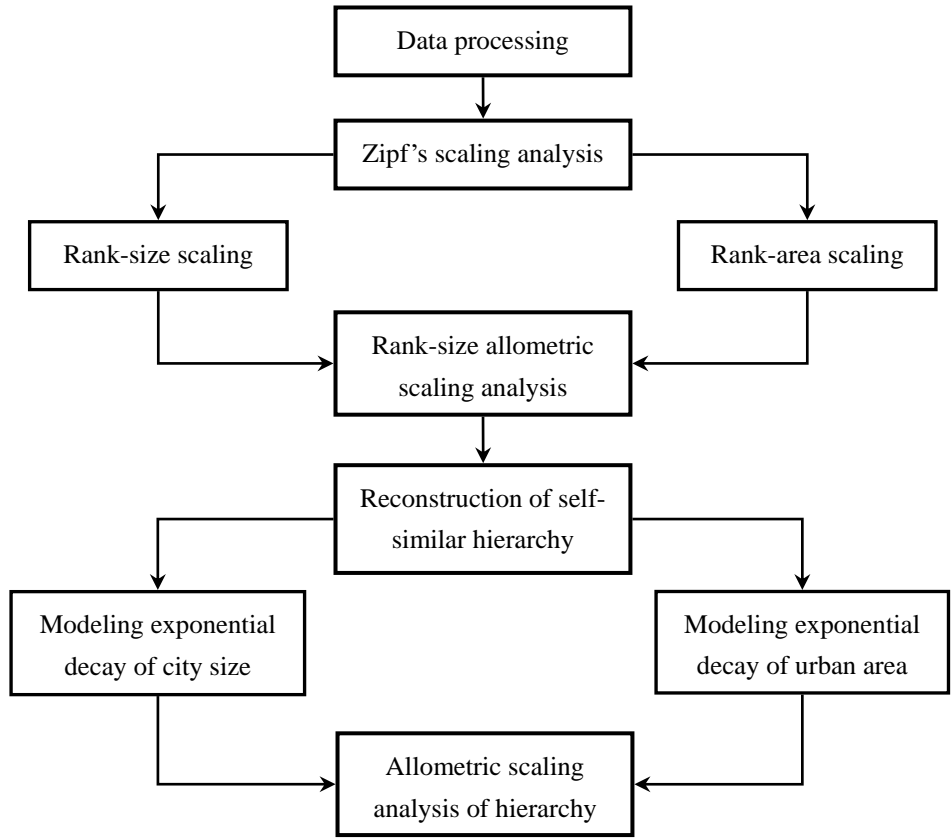

**Figure 9 The general process of hierarchical rescaling and allometric analyses of urban systems**

Compared with previous studies on the allometry and scaling of Chinese cities, this work bears significant characteristics. Firstly, the early studies are based on the two-parameter Zipf's law, but Chinese cities follow the three-parameter Zipf's law. Using hierarchical scaling to replace the rank-size scaling, we can catch the characters of the three-parameter Zipf distributions (Chen, 2016). Secondly, the early studies are based on the rank-size distribution instead of hierarchical structure of urban systems. This paper is based on datasets of cascade sequences abstracting from self-similar hierarchy of cities. In this instance, the allometric scaling can be revealed obviously. By reconstructing a hierarchy with cascade structure, we can reduce the random disturbance of noises in observational data. In particular, the rank-size patterns can be brought to light. The allometric scaling relation comes from Zipf's distributions of urban population and area. Allometric growth law can provide circumstantial evidence of Zipf's law. Maybe some persons prefer the common rank-size scaling to the hierarchical scaling analysis, because the traditional method is simple and the sample of cities seems to be bigger. As a matter of fact, the effect of statistical analysis depends on the association of degree of freedom with level of significance rather sample size. As indicated above, the common Zipf's law and allometric growth law are contained in the hierarchical rescaling process. Third, two approaches can be employed to estimate the allometric scaling exponents based on self-similar hierarchies. One is the OLS-based approach, and the other, the MLE-based approach. The former can be directly applied to this allometric scaling analysis, while the latter can only be indirectly applied to the hierarchical series data. The MLE-based method proposed by Clauset *et al* (2009) is designed for binned data rather than cascade sequences. If the allometric scaling relation of cities is well developed, the two approaches will lead to similar computational results. However, if the allometric scaling is not well developed, the OLS-based approach is better than the MLE-based approach. This suggests that the direct approach is better than the indirect approach.

The deficiencies of this study are as follows. First, the sample comprises the about 660 officially approved cities. In China, there are thousands of human settlements that can be treated as cities. However, only the observational data of the officially approved cities are available for quantitative studies. Perhaps just because of this, the one- or two-parameter Zipf's distribution was replaced by the three-parameter Zipf's distribution. Second, the data are based on statistical investigation rather than census. On the one hand, the census data are not continuous in time; on the other, we cannot find the land survey data for urbanized area. What is more, city population is not just within the limit of urban area. In fact, the datasets of natural data have better quality (Jiang and Jia, 2011; Jiang and Liu, 2012). These data have been employed to make allometric scaling analysis and the results are satisfying (Chen and Jiang, 2016). From the datasets of natural cities of America, England, France, and Germany, we can derive the same hierarchical allometric scaling as that of China. Unfortunately, we have no long sample path data of urban area and population from the time series of natural cities. Thus, only the cross-sectional analysis rather than the dynamic analysis can be made for the time being. Third, the hierarchy is constructed by city number rather than population size. There are two approaches to reconstructing a hierarchy of human settlements (Jiang and Yao, 2010): one is by size presented by Davis (1978), and the other is by number developed by Chen (2008). For simplicity, this study is only focused on number-based hierarchy. Fourth, the hierarchical allometric rescaling is based on population size rather than urban area. Urban population is not always consistent with urban area, that is, $P(k)$ greater than $P(k+1)$ does not necessarily indicate that $A(k)$ is greater than $A(k+1)$. We can rank cities either by population size or by urban area (Figure 10). There is subtle difference between the two results of hierarchies. Due to limited space, the area-based hierarchy is not discussed in this article. Despite all these shortcomings, the allometric scaling and its evolution patterns of Chinese cities are brought to light.

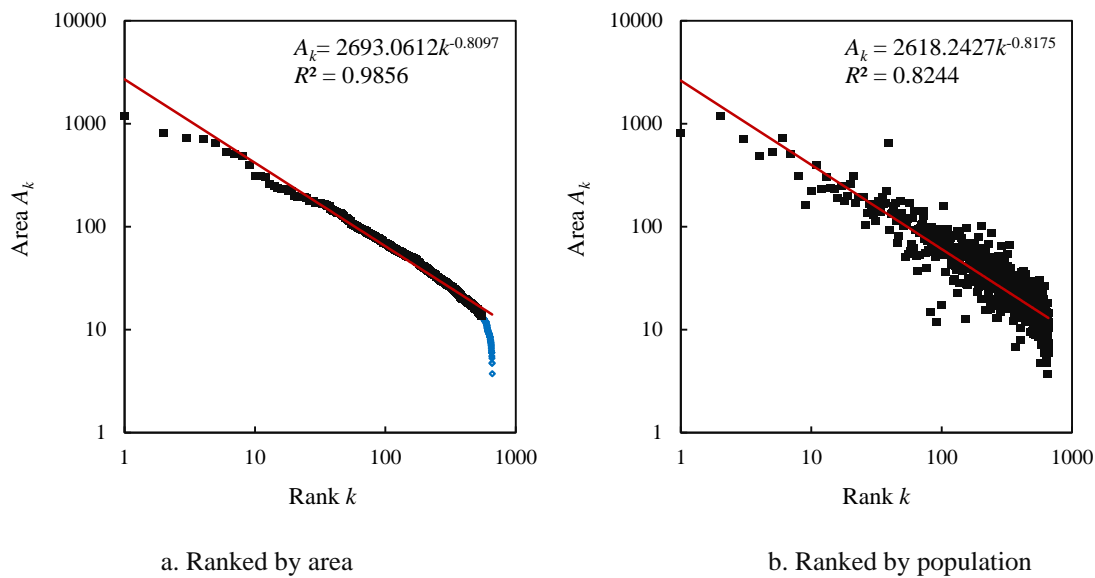

a. Ranked by area    b. Ranked by population

**Figure 10 The rank-area distributions of Chinese cities based on built-up area and population size in 2005**

## 5. Conclusions

A new analytical framework is presented by means of the case study of Chinese cities. In this framework, Zipf's law and allometric growth law are integrated into hierarchical scaling analysis of urban systems. Using the analytical process, we brought to light the spatio-temporal evolution properties of Chinese cities. By this study, we can get insight into the rank-size patterns and allometric scaling of Chinese system of cities. The main points of this work are as follows. **First, Chinese cities follow allometric scaling law that is based on Zipf's law. However, the rank-size pattern and allometric scaling emerged from the nonlinear dynamics of urbanization.** On the whole, the hierarchy of Chinese cities follows the allometric scaling law. However, from 1998 to 2000, the power-law relation degenerated to a linear relation. From 2003 on, the rank-size scaling and allometric scaling become clearer and more significant over time. This lends further support to the suggestion that both the allometric scaling and the rank-size scaling are evolutional laws. **Second, the allometric scaling exponent reflects the characters of the rank-size distribution and the** **man-land relation of Chinese cities.** Where the global level based on all the cities is concerned, the allometric scaling exponent values were relatively stable and approached to the theoretically expected value 0.85. However, where the local level based on the cities within the scaling range, the allometric scaling exponent values departed from 0.85 and approached 0.93. Because of fast urbanization and real estate industry, the allometric exponent values took on a rising trend for a time. This implies that the *per capita* land use area in Chinese cities does not fall but rise due to the increase of city sizes. In fact, the real estate bubble economy results in the waste of land resource utilization in China. After 2009, the local allometric exponent trended down to 0.85. **Third, the allometric scaling of Chinese cities is associated with the Zipf's distributions of urban population and area.** If and only if the urban population size and urban area follow Zipf's law, the allometric scaling relation between urban area and population emerge. The rank-size allometry can be converted into hierarchical allometry of urban systems. Before 2003, the allometric scaling relation degenerated to a linear relation on one occasion owing to the city population size failed to comply with Zipf's law. In this studies, the linear relationships are technically treated as power-law relations. From 2003 onwards, both the urban population and area conform to Zipf's law, and thus the rank-size allometry and hierarchical allometry emerged from the process of urbanization. Both the OLS-based results and MLE-based results supports this conclusion meanwhile.

## Acknowledgements

This research was sponsored by the National Natural Science Foundation of China (Grant No. 41590843 & 41671167). The supports are gratefully acknowledged.